\documentstyle[twocolumn,floats,prl,aps]{revtex}
\tighten

\def\srs{\scriptscriptstyle}
\def\srr{\rm\scriptscriptstyle}
\def\arraylinestretch{8pt}
\def\arraylinestretchlarge{15pt}
\newcommand{\dif}[2]{\protect\frac{\protect\partial{}#1}{\protect
\partial{}#2}}
\newcommand{\ddif}[2]{\protect\frac{{\rm d}#1}{{\rm d}#2}}
\newcommand{\df}{{\rm d}}
\newcommand{\im}{{\rm i}}
\newcommand{\beq}{\begin{equation}}
\newcommand{\enq}{\end{equation}}
\newcommand{\beqarr}{\begin{eqnarray}}
\newcommand{\enqarr}{\end{eqnarray}}
\newcommand{\ex}{{\rm \/e\/}}

\newcommand{\RE}{\mathop{\rm Re}\nolimits}
\newcommand{\nmbr}[1]{{\rm\mathcode`\,="2C #1}}
\newcommand{\eh}[1]{{\rm #1}}

\begin{document}

\title{Quantile motion of electromagnetic waves in wave 
guides of varying cross section and dispersive media}
\author{H.D. Dahmen, E. Gjonaj, T. Stroh}
\address{Fachbereich Physik, Universit\"at Siegen, 
57068 Siegen, Germany}
\date{\today}
\maketitle

\begin{abstract}
We discuss applications of the quantile concept of trajectories
and velocities to the propagation of electromagnetic signals in
wave guides of varying cross section. Quantile motion is a
general description of the transport properties of measurable
conserved quantities in quantum mechanics as well as in
classical electrodynamics. In the latter case we consider the
quantile motion of electromagnetic energy as the direct result
of a physical measurement.  In this sense the quantile velocity
corresponds to the electromagnetic signal velocity also in the
presence of barriers and inhomogeneities in the medium of
propagation. We show that this signal velocity is always smaller
than the speed of light in vacuum. Using numerical examples we
demonstrate how typical wave phenomena can be described in terms
of the quantile motion.
\end{abstract} 


\section{Introduction}
In recent years the tunneling of an evanescent electromagnetic
pulse in a wave guide with a cross section reduced over part of
the longitudinal extension of the wave guide has been studied. 
A series of microwave experiments \cite{bi-1,bi-2,bi-3,bi-4}
claims to have found superluminal tunneling velocities. The
velocity of the signal has been identified with the velocity 
of the maximum of the pulse.

A physical detector always needs a certain amount of energy 
to change its state, therefore we use the quantile velocity 
of a density obtained by way of normalizing the electromagnetic
energy density of the pulse to unity \cite{bi-5,bi-6,bi-7}.

A numerical simulation of the microwave experiments with wave
packets in wave guides with varying cross section is carried
through. The quantile velocity of wave packets is studied
numerically. It is shown rigorously that the quantile velocity
is not superluminal. A short discussion of the quantile velocity
in dispersive and absorptive media is presented. For a critical
discussion of superluminal velocities of electromagnetic signals
see also P. Thoma at al. \cite{bi-8}, G. Raithel \cite{bi-9}, G.
Diener \cite{bi-10}, and the contribution of H. Goenner in this
workshop \cite{bi-11}.

\section{Stationary waves in wave guides of varying cross 
sections. Unitarity relations}
We shall consider a wave guide of rectangular cross section 
and perfectly conducting walls, extending in the longitudinal
$z$ direction from $-\infty$ to $+\infty$. The simplest
construction allowing an interpretation in terms of tunneling
for an electromagnetic signal propagating in the wave guide is
obtained by assuming a narrowing of the cross section, e.g., in
the transverse $x$ direction extending to the left of the origin
for $z>0$ as shown in Fig.~\ref{fig-1}. An electromagnetic wave
moving from the right towards the narrowing of the cross section
will be affected in a similar way as the quantum-mechanical wave
function of a particle incident onto a potential step at 
$z = 0$.

To further simplify the problem we consider only the propagation
of {\sl TE}-waves of the type ${\bf H}_{n0}$ for $n = 1,2,\dots$.
Such a situation can indeed be realized if we choose a constant
height of the wave guide ($b$ in Fig.~\ref{fig-1}) small enough
for all waves of the type ${\bf H}_{nm}$ with $m \geq 1$ to be
evanescent \cite{bi-12}.  Thus, the stationary electromagnetic
field in the wave guide reduces to three field components,
$E_y$, $H_x$, and $H_z$, each depending on $x$ and $z$.  One of
the field components can be considered as independent, e.g.,
$E_y$, whereas the other two follow from the Maxwell equations,
e.g., $\im\omega\mu\mu_{0}{\bf H} = \nabla\times {\bf E}$, where
$\omega$ is the frequency of the wave and $\mu = 1$ is the
magnetic permeability of vacuum. It is worth noting that
according to a well-known result \cite{bi-13,bi-14} the magnetic 
field components are singular at the sharp edge of the
narrowing. This implies an increase of the magnetic energy 
at the cost of the electric energy in the wave guide
\cite{bi-12,bi-14}.

The solution for the stationary electric field component $E_y$
fulfilling the Helmholtz wave equation with the appropriate
boundary conditions is a superposition of modes
\catcode`@=11
\if@twocolumn
\beq
E_{y}^{n0}(x,z) = \left\{\begin{array}{ll}
\sin(\frac{n\pi x}{a})\ex^{\pm \im\kappa_{n}z} \: , 
\quad & z < 0 \: ,
\\[\arraylinestretch]
\sin(\frac{n \pi x}{a'})\ex^{\pm \im\kappa_{n}'z} \: , 
\quad & z > 0 \: ,
\end{array}
\right.
\label{eq-1}
\enq
where we denote $\kappa_{n} = \sqrt{(\omega/c)^{2} - (n\pi/
a)^{2}}$ and $\kappa_{n}' = \sqrt{(\omega/c)^{2} - (n\pi/
a')^{2}}$ the discrete wave numbers associated with the two 
regions of the wave guide. 
\else
\beq
E_{y}^{n0}(x,z) = \left\{\begin{array}{lll}\displaystyle
\sin(\frac{n\pi x}{a})\ex^{\pm \im\kappa_{n}z} \: , \quad
& \displaystyle \kappa_{n} = \sqrt{(\frac{\omega}{c})^{2} -
(\frac{n\pi}{a})^{2}} \: , \quad & \displaystyle z < 0 \: ,
\\[\arraylinestretchlarge]\displaystyle 
\sin(\frac{n \pi x}{a'})\ex^{\pm \im\kappa_{n}'z} \: , \quad
& \displaystyle \kappa_{n}'= \sqrt{(\frac{\omega}{c})^{2} -
(\frac{n\pi}{a'})^{2}} \: , \quad & \displaystyle z > 0 \: ,
\end{array}\right.
\label{eq-1}
\enq
where $\kappa_{n}$ and $\kappa_{n}'$ ($n = 1,2,\dots$) are
discrete wave numbers associated with the two regions of the
wave guide.
\fi
Considering a given mode ${\bf H}_{m0}$ as the incoming wave
incident from the right ($z < 0$) we are led to the ansatz 
for the solution
\if@twocolumn
\beq
E_y(x,z) = \left\{
\begin{array}{l}
\sin(\frac{m\pi x}{a}) \ex^{\im\kappa_{m}z}
+ \sum_{n=1}^{\infty} {\cal A}_{n}^{\srr R}
\sin(\frac{n\pi x}{a}) \ex^{-\im\kappa_{n}z} 
\\[\arraylinestretch]
\sum_{n=1}^{\infty} {\cal A}_{n}^{\srr T}
\sin(\frac{n\pi x}{a'}) \ex^{\im\kappa'_{n}z} 
\end{array}\right.
\label{eq-2}
\enq
for $z < 0$ and $z > 0$, respectively.
\else
\beq
E_y(x,z) = \left\{\begin{array}{ll}\displaystyle
\sin(\frac{m\pi x}{a}) \ex^{\im\kappa_{m}z}
+ \sum_{n=1}^{\infty} {\cal A}_{n}^{\srr R}
\sin(\frac{n\pi x}{a}) \ex^{-\im\kappa_{n}z} \: ,
\quad & \displaystyle z < 0 \: ,
\\[\arraylinestretchlarge]\displaystyle
\sum_{n=1}^{\infty} {\cal A}_{n}^{\srr T}
\sin(\frac{n\pi x}{a'}) \ex^{\im\kappa'_{n}z} \: , 
& \displaystyle z > 0 \: .
\end{array}\right.
\label{eq-2}
\enq
\fi
Here, ${\cal A}_{n}^{\srr R}$ and ${\cal A}_{n}^{\srr T}$ are
the (complex) amplitudes of the reflected and transmitted modes
in complete analogy to the one-dimensional tunneling in quantum
mechanics.  However, an infinite number of different modes is
needed in order to fulfill the boundary conditions at $z = 0$, 
so that the type of the outgoing wave does not necessarily 
coincide with that of the incoming wave.

Requiring continuity for the electric field component $E_y$ 
at the wave-guide aperture at $z = 0$, we obtain for the 
amplitudes ${\cal A}_{n}^{\srr R}$ and ${\cal A}_{n}^{\srr T}$
\if@twocolumn
\beq
\left\{\begin{array}{l}
\delta_{mn} + {\cal A}_{n}^{\srr R} = 
\frac{2}{a}\int_{0}^{a'}\df x \:
\epsilon(x)\sin(\frac{n\pi x}{a}) \: , 
\\[\arraylinestretch]
{\cal A}_{n}^{\srr T} = 
\frac{2}{a'}\int_{0}^{a'} \df x \:
\epsilon(x)\sin(\frac{n\pi x}{a'})  \: ,
\end{array}\right.
\label{eq-3}
\enq
\else
\beq
\delta_{mn} + {\cal A}_{n}^{\srr R} = 
\frac{2}{a}\int_{0}^{a'}\df x \:
\epsilon(x)\sin(\frac{n\pi x}{a}) \: ,\qquad
{\cal A}_{n}^{\srr T} = 
\frac{2}{a'}\int_{0}^{a'} \df x \:
\epsilon(x)\sin(\frac{n\pi x}{a'})  \: ,
\label{eq-3}
\enq
\fi
where $n = 1,2,\dots$ and $\epsilon(x) = E_y(x,z = 0)$ is 
the value of the electric field strength at the aperture. 
Denoting $\delta_{mn} + {\cal A}_{n}^{\srr R}\equiv{\cal
A}_{n}^{\srs(1)}$ and ${\cal A}_{n}^{\srr T}\equiv{\cal
A}_{n}^{\srs (2)}$ we can equivalently write equations
(\ref{eq-3}) as
\if@twocolumn
\beq
{\cal A}_{n}^{\srs(1)}=\frac{2}{a}\sum_{k=1}^{\infty}
\Lambda_{nk}{\cal A}_{k}^{\srs (2)} \: ,
\label{eq-4}
\enq
where
\[
\Lambda_{nk}=(-1)^k\frac{a'k}{\pi}
\frac{\sin(n\pi a'/a)}{(na'/a)^2-k^2} \: .
\]
\else
\beq
{\cal A}_{n}^{\srs(1)}=\frac{2}{a}\sum_{k=1}^{\infty}
\Lambda_{nk}{\cal A}_{k}^{\srs (2)} \: ,
\quad
\Lambda_{nk}=(-1)^k\frac{a'k}{\pi}
\frac{\sin(n\pi a'/a)}{(na'/a)^2-k^2} \: .
\label{eq-4}
\enq
\fi

The continuity condition for the transverse magnetic field
component at $z = 0$ yields
\if@twocolumn
\beq
\begin{array}{c}
\sum_{n=1}^{\infty}\kappa_{n}'{\cal A}_{n}^{\srs (2)}
\sin(\frac{n\pi x}{a'}) = 2\kappa_{m}\sin(\frac{m\pi x}{a})
\\[\arraylinestretch]
-\sum_{n=1}^{\infty}\kappa_{n}{\cal A}_{n}^{\srs(1)}
\sin(\frac{n\pi x}{a}) \: .
\end{array}
\label{eq-5}
\enq
\else
\beq
\sum_{n=1}^{\infty}\kappa_{n}'{\cal A}_{n}^{\srs (2)}
\sin(\frac{n\pi x}{a'}) = 2\kappa_{m}\sin(\frac{m\pi x}{a})
- \sum_{n=1}^{\infty}\kappa_{n}{\cal A}_{n}^{\srs(1)}
\sin(\frac{n\pi x}{a}) \: .
\label{eq-5}
\enq
\fi
Introducing equations (\ref{eq-3}) into (\ref{eq-5}) we 
obtain for the electric field $\epsilon(x)$ at the aperture 
the integral equation
\if@twocolumn
\beq
\int_{0}^{a'}\df x'\: {\cal K}(x,x')\epsilon(x') = \kappa_{m}
\sin(\frac{m\pi x}{a})
\label{eq-6}
\enq
for $0\leq x\leq a$. The kernel ${\cal K}(x,x')$ is given by
\else
\beq
\int_{0}^{a'}\df x'\: {\cal K}(x,x')\epsilon(x') = \kappa_{m}
\sin(\frac{m\pi x}{a})\: , \quad 0\leq x\leq a \: ,
\label{eq-6}
\enq
where the kernel ${\cal K}(x,x')$ is given by
\fi
\if@twocolumn
\beq
\begin{array}{c}
{\cal K}(x,x') = \sum_{n=1}^{\infty}\left[\frac{\kappa_{n}}{a}
\sin(\frac{n\pi x}{a})\sin(\frac{n\pi x'}{a})\right.
\\[\arraylinestretch]
\left. + \frac{\kappa'_{n}}{a'}\sin(\frac{n\pi x}{a'})
\sin(\frac{n\pi x'}{a'})\right] \: .
\end{array}
\label{eq-7}
\enq
\else
\beq
{\cal K}(x,x') = \sum_{n=1}^{\infty}\left[\frac{\kappa_{n}}{a}
\sin(\frac{n\pi x}{a})\sin(\frac{n\pi x'}{a}) +
\frac{\kappa'_{n}}{a'}\sin(\frac{n\pi x}{a'})
\sin(\frac{n\pi x'}{a'})\right] \: .
\label{eq-7}
\enq
\fi
The reflection and transmission amplitudes ${\cal A}_{n}^{\srr
R}$ and ${\cal A}_{n}^{\srr T}$ follow then from the equations
(\ref{eq-3}) and the solution of the integral equation
(\ref{eq-6}).

In general the integral equation (\ref{eq-6}) is not 
solvable in closed form and the field $\epsilon(x)$ has to be 
determined numerically. We can, however, try to find approximate
expressions for ${\cal A}_{n}^{\srr R}$ and ${\cal A}_{n}^{\srr
T}$ by using the alternative representation of the integral
equation (\ref{eq-6}) as a system of linear equations for the
coefficients ${\cal A}_{n}^{\srs (2)}$,
\if@twocolumn
\beq
\sum_{{n}=1}^{\infty}\left[\delta_{nk} + 
\frac{4}{aa'\kappa'_{k}}T_{nk}\right]
{\cal A}_{n}^{\srs (2)} = 
\frac{4\kappa_{m}}{a'\kappa'_{k}}\Lambda_{mk} 
\label{eq-8}
\enq
with ${k} = 1,2,\dots$ and matrix elements $T_{nk}$ given by
\else
\beq
\sum_{{n}=1}^{\infty}\left[\delta_{nk} +
\frac{4}{aa'\kappa'_{k}}T_{nk}\right]
{\cal A}_{n}^{\srs (2)} = 
\frac{4\kappa_{m}}{a'\kappa'_{k}}\Lambda_{mk} \: ,
\qquad {k}=1,2,\dots \: ,
\label{eq-8}
\enq
where the matrix elements $T_{nk}$ are given by
\fi
\beq
T_{nk} = \sum_{l=1}^{\infty}\kappa_{l}
\Lambda_{ln}\Lambda_{lk} \: .
\label{eq-9}
\enq
Under certain conditions discussed below we can assume the
matrix ${\bf T}$ to be nearly diagonal and put its elements 
in the form
\beq
T_{nk} = \kappa'_{k}(\frac{a'}{2})^2
\left(\delta_{nk}+m_{nk}\right) \: , 
\label{eq-10}
\enq
where ${\bf m}$ is a complex matrix with elements fulfilling
$|m_{nk}| \ll 1$ for $n,k = 1,2,\dots$. The solution of
(\ref{eq-8}) follows then from the expansion of the 
corresponding inverse matrix into a fast converging 
Neumann series
\beq
{\cal A}_{n}^{\srs (2)} = \frac{4\kappa_{m}}{a'}(\frac{a}{a+a'})
\sum_{k=1}^{\infty}\left[{\bf 1} - (\frac{a'}{a+a'}){\bf m} +
\cdots \right]_{nk} \frac{\Lambda_{mk}}{\kappa'_{k}} \: .
\label{eq-11}
\enq
Neglecting all the terms in (\ref{eq-11}) but the leading one 
we obtain the coefficients
\if@twocolumn
\beq
{\cal A}_{n}^{\srs (2)} = (-1)^n(\frac{a}{a+a'})
\frac{4n\kappa_{m}}{\pi\kappa'_{n}}
\frac{\sin(m\pi a'/a)}{(ma'/a)^2-n^2}
\label{eq-12}
\enq
for $n = 1,2,\dots$,
\else
\beq
{\cal A}_{n}^{\srs (2)} = (-1)^n(\frac{a}{a+a'})
\frac{4n\kappa_{m}}{\pi\kappa'_{n}}
\frac{\sin(m\pi a'/a)}{(ma'/a)^2-n^2} \: ,
\quad {n} = 1,2,\dots \: ,
\label{eq-12}
\enq
\fi
implying at $z = 0$ a transverse magnetic field component 
$H_x$ given by
\beq
H_x(x,0)=-(\frac{2a}{a+a'})\frac{\kappa_{m}}{\omega\mu\mu_0}
\sin(\frac{m\pi x}{a}) \: .
\label{eq-13}
\enq
Equation (\ref{eq-13}) predicts the increase in the magnitude 
of the transverse magnetic field component $H_x$ at $z = 0$,
whereas the shape of the field there obviously coincides with
that of the incoming mode ${\bf H}_{m0}$. Thus, the
approximation leading to (\ref{eq-12}) seems appropriate in the
limiting case of geometric optics, i.e., if the wave length of
the incoming wave is short compared to the cross sections $a$
and $a'$ of the wave guide. In the case of an evanescent wave for
$z > 0$, however, more terms in the expansion (\ref{eq-11}) are
needed for the boundary conditions at $z = 0$ to be fulfilled.
Therefore, a numerical solution of equation (\ref{eq-6}) is used
in what follows.

We now turn to a more complicated geometry for the wave guide
involving two boundary conditions at $z = 0$ and $z = L$, where
$L$ is the length of a symmetrically placed barrier (reduction
of the cross section as shown in Fig.~\ref{fig-2}). Numerical
simulations for the stationary field components in the wave
guide are shown in Fig.~\ref{fig-3} and Fig.~\ref{fig-4}. The
continuity of the transverse field components at the wave-guide
aperture as well as the magnetic-field singularities at the
narrowing edges are shown in Fig.~\ref{fig-3}. Figure~\ref{fig-4}
is a spectral diagram with reflection and transmission
coefficients (in the regions $z < 0$ and $z > L$, respectively)
of the incoming as well as of higher modes.

Because of the magnetic-field singularities in a real experiment
the maxima of the magnetic field strength would always be located 
in the vicinity of the wave-guide edges. Therefore, they give no 
information on the velocity of a tunneling electromagnetic signal.
If, instead, we refer to the electromagnetic energy density in the 
wave guide we can avoid dealing with singularities by recalling 
that the singularities of the energy density are integrable. 
Poynting's theorem in our case reads
\beq
\dif{w_{\srr em}(x,z,t)}{t} + \dif{s_x(x,z,t)}{x}
+ \dif{s_z(x,z,t)}{z} = 0 \: ,
\label{eq-14}
\enq
where $w_{\srr em}$ is the two-dimensional energy density and
$s_x$ and $s_z$ are the components of the Poynting vector.
Integrating (\ref{eq-14}) over the transverse direction $x$ 
the second term in (\ref{eq-14}) vanishes and we obtain
\beq
\dif{}{t}{\cal W}(z,t) + \dif{}{z}{\cal S}_z(z,t) = 0\: ,
\label{eq-15}
\enq
where ${\cal W}(z,t)$ and ${\cal S}_z(z,t)$ are now the
one-dimensional energy and current densities in the longitudinal
direction $z$. These quantities are free of singularities and
fulfill the one-dimensional continuity equation (\ref{eq-15}) in
complete analogy to the probability and current density in the
one-dimensional tunneling in quantum mechanics \cite{bi-15}.
Therefore, it is appropriate to consider the energy density
${\cal W}(z,t)$ instead of the electric and magnetic field
strengths in order to investigate the tunneling properties 
of electromagnetic signals in wave guides.

Returning to the stationary fields, we may use (\ref{eq-15}) 
to obtain unitarity relations between the reflection and
transmission coefficients of the modes ${\cal A}_{n}^{\srr R}$
and ${\cal A}_{n}^{\srr T}$, $n = 1,2,\dots$. We find the
time-averaged longitudinal current
\beq
\widetilde{\cal S}_z(z) = -\frac{1}{2}\RE\int
\df x\: E_y(x,z) H_x^{*}(x,z)
\label{eq-16}
\enq
to be a constant along the wave guide. Comparing the expressions
for $\widetilde{\cal S}_z(z)$ in the reflection and transmission
region we obtain
\beq
\sum_{n=1}^{n_{\rm c}}\frac{\kappa_{n}}{\kappa_{m}}
|{\cal A}_{n}^{\srr R}|^2 + \frac{a'}{a}\sum_{n=1}^{n'_{\rm c}}
\frac{\kappa'_{n}}{\kappa_{m}}|{\cal A}_{n}^{\srr T}|^2 = 1 \: ,
\label{eq-17}
\enq
where ${\bf H}_{m0}$ is the incoming mode, ${\bf H}_{n_{\rm c}
0}$ and ${\bf H}_{n_{\rm c}' 0}$ are the cutoff modes in the
reflection and transmission region, respectively. The appearance
in (\ref{eq-17}) of the upper limits $n_{\rm c}$ and $n_{\rm
c}'$ in the summation index $n$ is a consequence of time
averaging the current density ${\cal S}_z(z)$.  Thus, equation
(\ref{eq-17}) does not imply that only propagating modes $n \leq
n_{\rm c}$ ($n \leq n_{\rm c}'$) are responsible for the energy
transport through the barrier. In the microwave experiments
\cite{bi-1,bi-2,bi-3,bi-4} the cross sections $a$ and $a'$ of
the reflection and transmission region were the same. With this
assumption equation (\ref{eq-17}) reads
\beq
\sum_{n=1}^{n_{\rm c}} \frac{\kappa_{n}}{\kappa_{m}}
\left(|{\cal A}_{n}^{\srr R}|^2 + |{\cal A}_{n}^{\srr T}|^2
\right) = 1 \: .
\label{eq-18}
\enq
Further simplifying the situation by allowing the ground mode
${\bf H}_{10}$ as the only propagating mode in the wave guide
yields the unitarity relation of the form
\beq
|{\cal A}_{1}^{\srr R}|^2+|{\cal A}_{1}^{\srr T}|^2 = 1
\label{eq-19}
\enq
which clearly coincides with the corresponding equation in 
the one-dimensional quantum-mechanical tunneling.

\section{Tunneling of wave packets. Comparison with experiments}
In the following we construct tunneling wave packets in the 
wave guide as a superposition of the stationary solutions found
above. For such a superposition to be a propagating, incoming
wave in the region to the right of the barrier ($z < 0$) it 
must not contain wave components below the cutoff frequency
$\omega_{\rm c}$ (evanescent components) in this region.
Therefore we use, e.g., for the electric field component 
$E_y$ the expression
\beq
E_y(x,z,t)=\int_{\omega_{\rm c}}^{\infty} \df \omega \: 
f(\omega) E_{y {\rm s}}(x,z;\omega) \ex^{-\im\omega t} \: ,
\label{eq-20}
\enq
where $f(\omega)$ is the spectral distribution of frequencies
normalized in the interval $(\omega_{\rm c}, \infty)$ and 
$E_{y {\rm s}}$ is the stationary electric field component.

In Fig.~\ref{fig-5}--Fig.~\ref{fig-7} numerical simulations 
with Gaussian-like wave packets in the wave guide are shown. 
The spectral distribution $f(\omega)$ in this case is given by
\beq
f(\omega) = \frac{1}{\tilde{N}} \frac{\Theta(\omega 
- \omega_{\rm c})}{(\sqrt{2\pi}\sigma_\omega)^{1/2}}
\exp\left(-\frac{(\omega - \omega_0)^2}{4\sigma_\omega^2}
\right) \: ,
\label{eq-21}
\enq
where $\tilde{N}$ is a normalization factor, $\omega_0$ is the
mean frequency and $\sigma_\omega$ is the spectral width of the
wave packet.

Figure~\ref{fig-5} demonstrates the tunneling process 
through a symmetrically placed barrier in the wave guide 
for an incoming mode of the type ${\bf H}_{10}$. Most of 
the frequencies in the spectrum (\ref{eq-21}) are taken above 
the cutoff frequency $\omega_{\rm c}'$ of the barrier region 
$L > z > 0$.  Correspondingly, the  transmission rate is high. 
We observe that the shape of the reflected and transmitted wave 
packets is substantially deformed. Among other reasons this 
is due to the energy transfer to higher modes other than the 
incoming mode ${\bf H}_{10}$. In terms of the evolution of 
the wave-packet maxima, we start with a single maximum in the 
incoming wave and end up with many transmitted maxima
propagating with different  velocities in the wave guide.

In Fig.~\ref{fig-6} the resonant tunneling of a Gaussian 
wave packet is shown. The frequency spectrum contains many
transmission resonances (see also Fig.~\ref{fig-4}) leading 
to multiple reflections of the wave packet at the barrier walls.

In Fig.~\ref{fig-7} we compute the longitudinal energy density
${\cal W} (z,t)$ in the time domain at a fixed position $z$
behind the barrier and for different barrier lengths $L$. The
wave packet was chosen such that tunneling takes place mainly in
the evanescent regime. Under this condition we observe that the
maximum of the wave packet can appear behind the barrier earlier
than when moving in the free space with the vacuum speed of
light $c$. This behavior becomes more obvious for large $L$ as
the transmission rate decreases significantly.

As another example we consider the tunneling of Kaiser--Bessel
wave packets \cite{bi-16} with a limited and discrete frequency
spectrum in a given interval $[\omega_{-}, \omega_{+}]$. The
spectral distribution is given by
\beq
f(n)=\frac{I_0\left[\pi\alpha\sqrt{1 - (2n/N)^2}
\right]}{I_0(\pi\alpha)}\: ,\quad 0\leq|n|\leq\frac{N}{2}\: ,
\label{eq-22}
\enq
where $\alpha$ is the parameter characterizing the width of 
the distribution and $N$ is the number of the stationary-wave
components with frequencies $\omega_n = \omega_{-} + (n +
\frac{N}{2})(\omega_{+} - \omega_{-})/N$ for $n = -N/2,\dots,
-1,0,1\dots,N/2$. The distribution (\ref{eq-22}) ensures 
optimality of the wave-packet localization in the time domain
and was also used in the microwave experiments \cite{bi-1}. In
Fig.~\ref{fig-8} we compute the tunneling time of the maximum of
a Kaiser--Bessel wave packet between the beginning and end of
the barrier as a function of the barrier length $L$. We consider
wave packets tunneling in the propagating and evanescent regime
and compare in the respective case the tunneling velocity of the
maximum with the group velocity in the wave guide and the
velocity of light in vacuum $c$. In the evanescent case we
observe the tunneling velocity of the maximum of the wave 
packet for long barriers to be independent of the barrier 
length $L$. This behavior corresponds to the Hartman effect 
which is well-known from the quantum-mechanical tunneling 
\cite{bi-17,bi-18}. Thus, evanescent tunneling of the maximum
becomes highly superluminal. In the example of Fig.~\ref{fig-8}
we obtain a tunneling time $\tau_{\rm T} \approx
\nmbr{127.78}\,{\rm ps}$ at a barrier length $L = \nmbr{100}
\,{\rm mm}$ for the maximum of the wave packet. The corresponding
tunneling velocity is $v_{\rm T} = L/\tau_{\rm T} \approx
\nmbr{2.6}c$ in very good agreement with the experimental 
result given in \cite{bi-1}.

\section{Quantile motion of electromagnetic signals.
Causality of signal propagation}

We measure the arrival time of a signal with a detector placed at the
fixed position $z_{\srr P}$. We assume that the region in which the
energy of an electromagnetic pulse is essentially different from zero
is initially far away from the position $z_{\srr P}$ of the detector.
The detection of the electromagnetic signal requires the deposition of
a certain amount $W$ of energy in the detector to cause a change of its
state indicating the arrival of the signal. This is equivalent to the
condition
\beq
\int_{z_{\srr P}}^{\infty}\df z\: {\cal W}(z,t) = W 
\label{eq-23}
\enq
on the time $t$ of arrival of the signal. Repeated 
measurements at different positions $z_{{\srr P}1}, z_{{\srr
P}2},\dots$ yield arrival times $t_1, t_2,\dots$ corresponding
to these positions.  They are discrete points on the trajectory
$z_{\srr P} = z_{\srr P}(t)$ defined by requiring the condition
\beq
\int_{z_{\srr P}(t)}^{\infty}\df z\: {\cal W}(z,t) = W 
\label{eq-24}
\enq
to hold at all times $t$. If we call $W_0$ the total 
energy contained in the pulse then $P = W/W_0$ is the 
fraction of energy needed for detection and $\varrho(z,t) =
{\cal W}(z,t)/W_0$ is the normalized energy density. Equation
(\ref{eq-24}) can be put into the form
\beq
\int_{z_{\srr P}(t)}^{\infty}\df z\: \varrho(z,t) = P \: , 
\quad 0 < P < 1
\label{eq-25}
\enq
which is the same as Equation (11) of \cite{bi-5}. Therefore,
$z_{\srr P} = z_{\srr P}(t)$ is the quantile trajectory of the
electromagnetic signal. As to be expected it depends on the
fraction $P$, and thus on the sensitivity $W = PW_0$ of the
detector. The signal velocity is then given by
\beq
v_{\srr P}(t)=\ddif{z_{\srr P}(t)}{t} \: .
\label{eq-26}
\enq

Examples of quantile trajectories for tunneling Gaussian 
signals in wave guides of varying cross section are given in
Fig.~\ref{fig-9} and Fig.~\ref{fig-10}. Figure~\ref{fig-9}
shows that the presence of a barrier in the wave guide may 
only lead to a slower signal propagation at any give time $t$ 
and for every detector sensitivity $P$. Thus, no quantile 
signal velocity larger than the speed of light in vacuum $c$ 
is possible. Especially, in the evanescent tunneling regime 
(see Fig.~\ref{fig-9}c) the tunneling velocity is much 
smaller than $c$, whereas, as physically expected, most 
of the trajectories turn back to the reflection region.

The behavior of the quantile trajectories for different $P$
values reflects several properties of the tunneling process. 
In Fig.~\ref{fig-9}b the reflected and transmitted quantile
trajectories split into trajectory bunches propagating with
different velocities in the wave guide. They correspond to the
electromagnetic modes produced in the tunneling as described in
the previous section. This coincidence between the behavior of
quantile trajectories and typical wave phenomena in tunneling
can be observed also in resonant tunneling (see
Fig.~\ref{fig-10}).

The causality of quantile motion can explicitly be derived, 
for instance, in the case of tunneling {\sl TE}-waves. Using 
the definition (\ref{eq-24}) and the continuity relation
(\ref{eq-15}) we derive the quantile velocity (\ref{eq-24}),
c.f. Equation (13) in \cite{bi-5},
\beq
v_{\srr P}(t) = \ddif{z_{\rm q}(t)}{t} = 
\frac{{\cal S}[z_{\srr P}(t),t]}{{\cal W} 
[z_{\srr P}(t),t]} \: .
\label{eq-27}
\enq
Trajectories solving this equation have also been studied 
by Holland \cite{b-18a}. The modulus of the velocity field 
characterizing the differential equation (\ref{eq-27}) is
\if@twocolumn
\beq
\begin{array}{c}\displaystyle
|v(z,t)| = \frac{2c\left|\sum_{n=1}^{\infty}
\RE[cE_y^{n}] \RE[H_x^{n}]\right|}{\left|\sum_{n=1}^{\infty}
\RE[cE_y^{n}]^2 + \RE[H_x^{n}]^2 + \RE[H_z^{n}]^2
\right|}\leq c \: ,
\end{array}
\label{eq-28}
\enq
\else
\beq
|v(z,t)| = \left|\frac{{\cal S}(z,t)}{{\cal W}(z,t)}\right|
= c \frac{\displaystyle 2\left|\sum_{n=1}^{\infty}
\RE[cE_y^{n}] \RE[H_x^{n}]\right|}{\displaystyle\left|
\sum_{n=1}^{\infty} \RE[cE_y^{n}]^2 + \RE[H_x^{n}]^2
+ \RE[H_z^{n}]^2\right|} \leq c \: ,
\label{eq-28}
\enq
\fi
Thus, $v_{\rm q}(t)$ never exceeds the vacuum speed of light
$c$, i.e., the signal propagation described by the quantile
trajectory is causal.  This result is a general property of 
the quantile motion and holds independently of the type of 
the tunneling wave.

\section{Note on the quantile motion in dispersive and 
absorptive media}
It has been known for a long time that electromagnetic 
signal propagation in the spectral region of a dispersive 
medium characterized by anomalous dispersion and strong
absorption leads to superluminal phase and group velocity
\cite{bi-19,bi-20,bi-21,bi-22}. Even though the shape of the
propagating signal may be substantially deformed in comparison
to the shape of the incoming wave, a propagation velocity has
been considered which coincides with the velocity of one of the
pulse maxima. Then the velocity obtained this way is again
superluminal \cite{bi-22}. This result was recently reconfirmed
by the photon experiments at Berkeley \cite{bi-23,bi-24} and by
the experiments with photonic barriers in wave guides 
\cite{bi-25}. However, if one considers instead of the pulse
maxima the energy transport in the medium no superluminal
velocities occur. Diener \cite{bi-10} suggests a procedure for
separating propagating and non-propagating parts of energy of 
the electromagnetic pulse and obtains subluminal velocities 
for the energy transport in the medium.

Because of the interaction of the electromagnetic field with 
the dispersive medium, energy is permanently transferred between 
the propagating pulse and the oscillating charges in the medium. 
In the presence of dissipation this exchange is characterized 
by a loss of energy, since part of the mechanical energy of 
the oscillators is steadily transformed into heat. Thus, 
three different kinds of energy have to be considered, the
electromagnetic energy of the pulse, the mechanical energy 
of the oscillating charges of the medium, and the thermal 
energy stored in the medium \cite{bi-26}. Each of them may 
be considered as the energy of an open subsystem of the closed
system characterized by the total energy which is conserved.

We apply the concept of quantile motion to the wave propagation
in dispersive media. Sufficient condition for this is the
existence of a measurable quantity with a positive spatial
density (see \cite{bi-5,bi-6,bi-7}).  Since the total energy in
the medium is conserved, its density fulfills a continuity
equation similar to (\ref{eq-14}). Thus, quantile trajectories
for the total energy in a dispersive and absorptive medium may
be easily defined in complete analogy to the above description
for the (conserved) electromagnetic energy in wave guides. The
application of quantile motion for the electromagnetic signal
propagation in dispersive media has been carried through in
\cite{bi-26}. In all cases it has been shown that the quantile
velocities for the signal remain below the vacuum speed of light
$c$. An argument analogous to the one leading to (\ref{eq-28})
shows rigorously that the quantile velocity in the oscillator
model of dispersive and absorptive media is always smaller than
the vacuum speed of light. This and related results will be
presented in a forthcoming publication.

\section{Concluding remarks}
The concept of quantile motion has been applied to the
propagation of electromagnetic waves in wave guides of varying
cross section and in dispersive and absorptive media. It has
been shown that the signal velocity measured with a detector of
finite sensitivity never becomes superluminal. In the context of
the propagation of electromagnetic signals the quantile velocity
is a generalization to detectors with finite sensitivity of
Sommerfeld's concept \cite{bi-19} of a front velocity describing
the speed of a signal measured with a detector of infinite
sensitivity.

\if@twocolumn
\def\boxdist{92.4}
\def\boxwidth{87.2}
\def\picdist{14}
\else
\def\boxdist{88}
\def\boxwidth{76.8}
\def\picdist{7}
\fi
\begin{figure*}
\unitlength=1mm
\noindent
\begin{picture}(85,42)
\thinlines
\put(0,0){\includegraphics{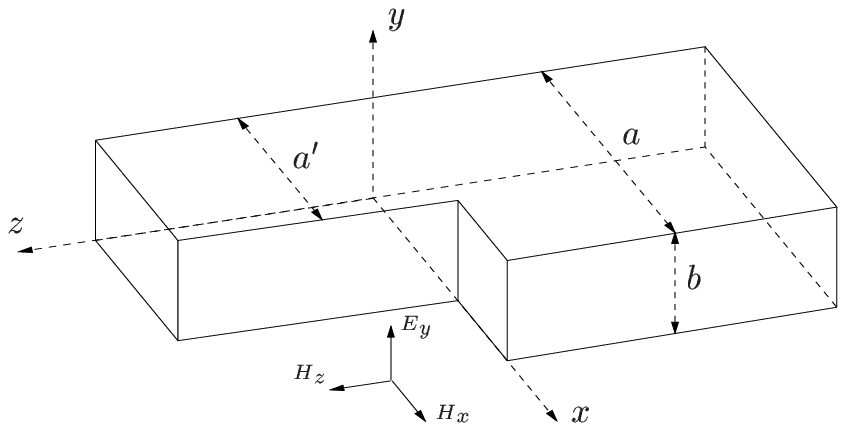}} 
\put(\boxdist,0){\makebox(\boxwidth,42)[b]
{\begin{minipage}[b]{\boxwidth mm}\protect\caption[]{Geometry of a 
wave guide with a reduction of the cross section beginning at 
$z = 0$ and extending to infinity. This configuration with an 
${\bf H}_{n0}$ wave incident from the right corresponds to a 
quantum-mechanical potential step.}
\end{minipage}}}
\end{picture}
\label{fig-1}
\end{figure*}

\begin{figure*}
\unitlength=1mm
\noindent
\begin{picture}(85,42)
\put(0,0){\includegraphics{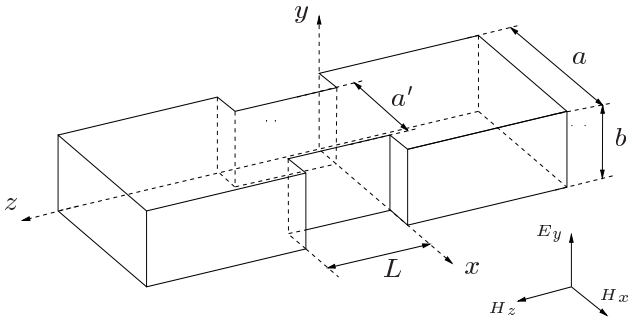}} 
\put(\boxdist,0){\makebox(\boxwidth,42)[b]
{\begin{minipage}[b]{\boxwidth mm}\protect\caption[]{Geometry of a 
wave guide with a symmetric reduction of the cross section 
between $z = 0$ and $z = L$ corresponding to a barrier of 
finite length $L$.}
\end{minipage}}}
\end{picture}
\label{fig-2}
\end{figure*}

\begin{figure*}
\unitlength=1mm
\noindent
\begin{picture}(85,120)
\put(0,0){\includegraphics{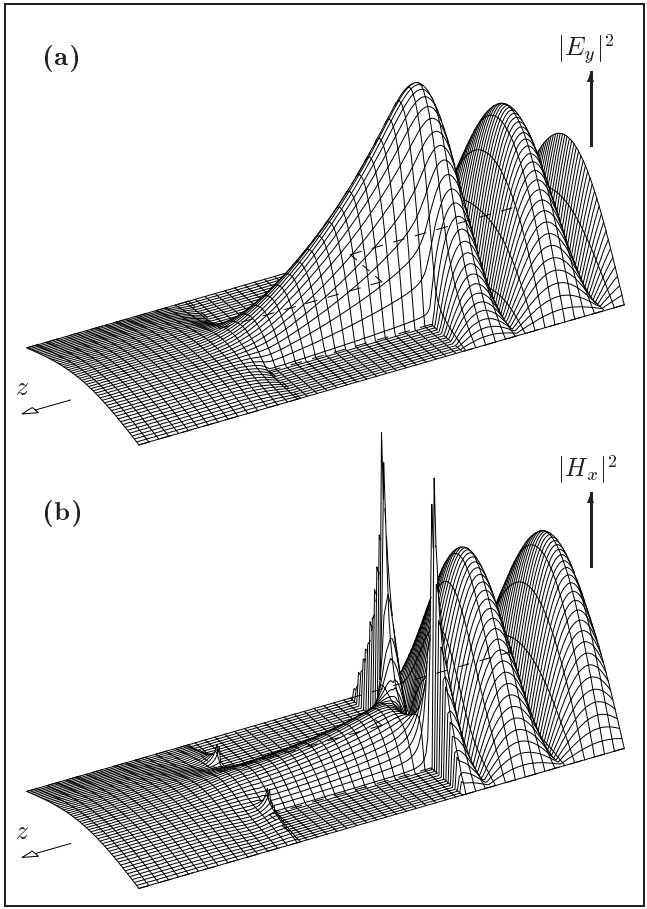}} 
\put(\boxdist,0){\makebox(\boxwidth,120)[b]
{\begin{minipage}[b]{\boxwidth mm}\protect\caption[]{Stationary 
fields in evanescent tunneling. The dimensions of the symmetric 
wave guide and of the barrier are $a = \nmbr{1}\,\eh{cm}$, 
$a' = \nmbr{0.4}\,\eh{cm}$, and $L = \nmbr{1}\,\eh{cm}$ (see 
Fig.~\ref{fig-2}). The incoming wave incident from the right 
is an ${\bf H}_{10}$ mode of frequency $\omega=\nmbr{21}\,
\eh{GHz}$. {\bf (a)} The electric field strength $|E_y|^2$ 
is shown. This component is continuous everywhere and vanishes 
at the walls of the wave guide.  {\bf (b)} The transverse 
magnetic field strength $|H_x|^2$ is shown. Obviously, the 
magnetic field is singular at the edges at $z = 0$ and 
$z = L$ of the wave guide.}
\end{minipage}}}
\end{picture}
\label{fig-3}
\end{figure*}

\begin{figure*}
\unitlength=1mm
\noindent
\begin{picture}(162,78)
\put(\picdist,0){\includegraphics{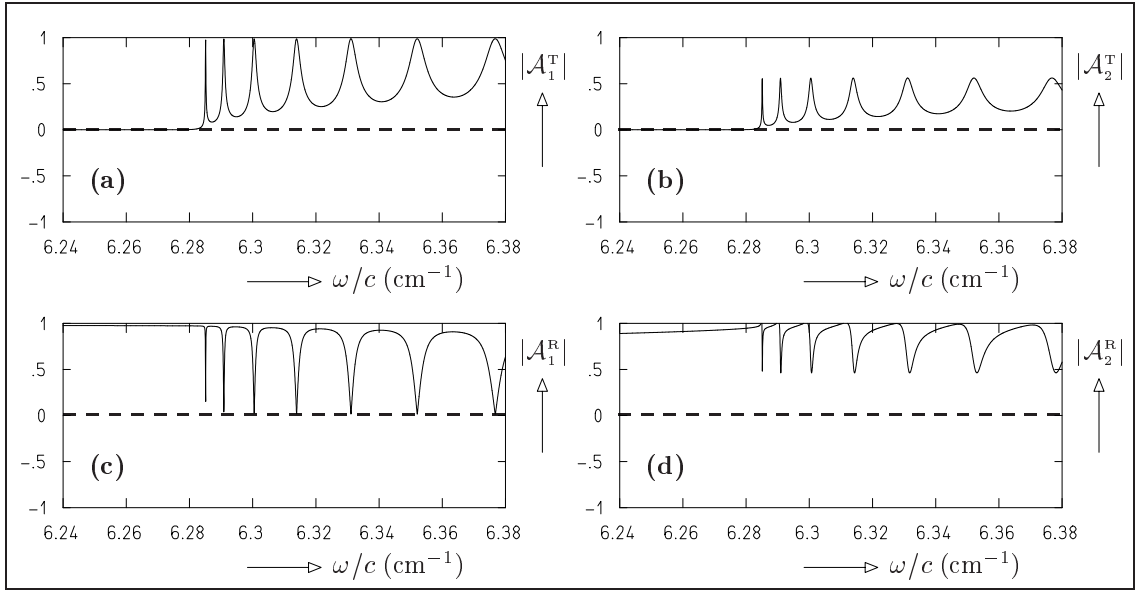}}
\end{picture}
\vspace*{0.2cm}
\caption[]{Spectral diagram with the transmission and 
reflection coefficients of the first first two modes 
${\bf H}_{10}$ and ${\bf H}_{20}$, where ${\bf H}_{10}$ 
is the incident mode. The dimensions of the wave guide 
and of the symmetric barrier are $a=\nmbr{1}\,\eh{cm}$,
$a'=\nmbr{0.5}\,\eh{cm}$, and $L=\nmbr{1}\,\eh{cm}$ (see
Fig.~\ref{fig-2}).}
\label{fig-4}
\end{figure*}

\begin{figure*}
\unitlength=1mm
\noindent
\begin{picture}(162,64)
\put(\picdist,0){\includegraphics{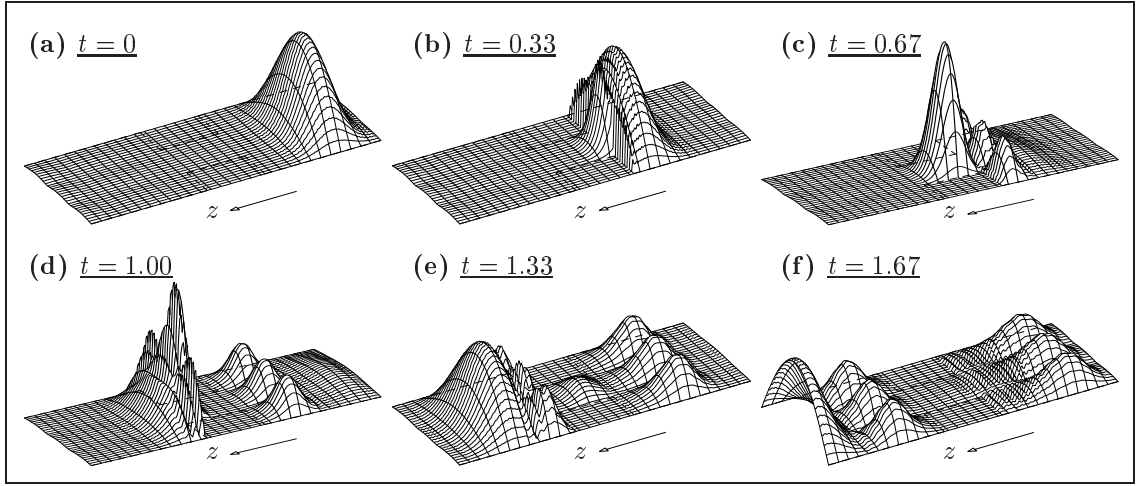}}
\end{picture}
\vspace*{0.2cm}
\caption[]{Time development of a tunneling Gaussian 
wave packet.  The electric field strength $|E_y|^2$ is shown.
The dimensions of the wave guide are $a=\nmbr{1}\,\eh{cm}$,
$a'=\nmbr{0.5}\,\eh{cm}$, and $L=\nmbr{5} \,\eh{cm}$. The incident
wave packet is of the type ${\bf H}_{10}$ with frequencies
centered at $\omega_0=\nmbr{30}\,\eh{GHz}$ and spectral width
$\sigma_\omega=\nmbr{1.5}\,\eh{GHz}$. The production of higher
modes in tunneling which propagate with different velocities 
in the wave guide, is observed. Times are given in picoseconds.}
\label{fig-5} 
\end{figure*}

\begin{figure*}
\unitlength=1mm
\noindent
\begin{picture}(162,98.5)
\put(\picdist,0){\includegraphics{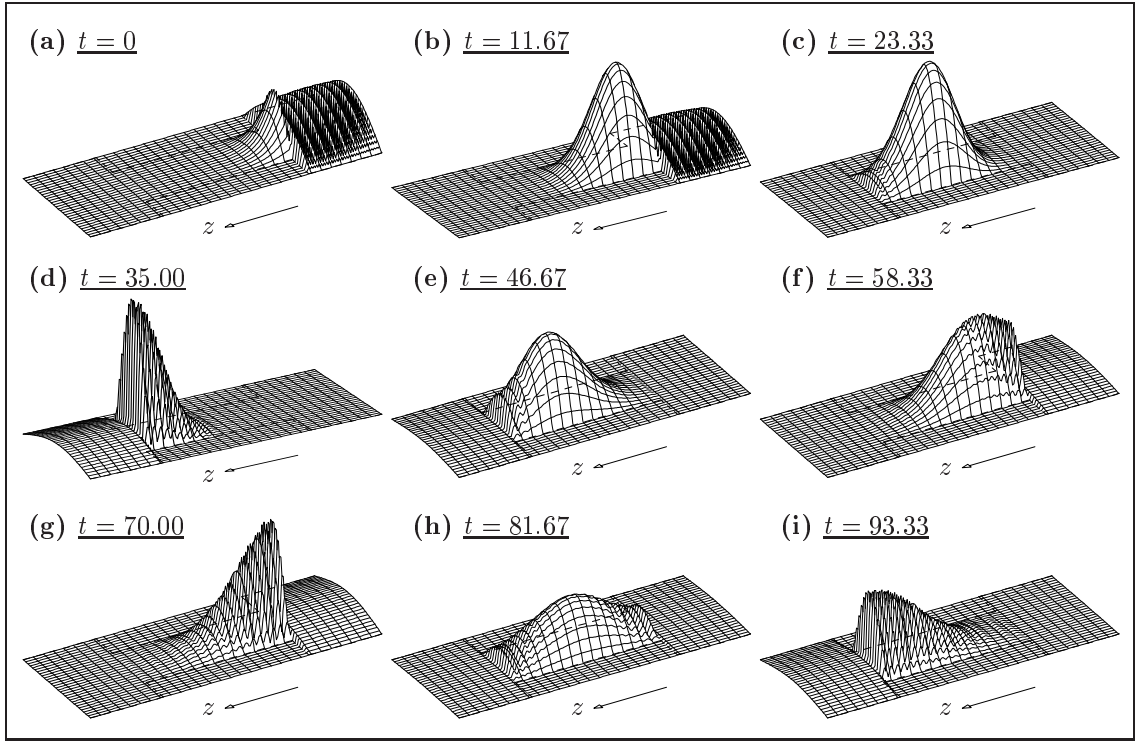}}
\end{picture}
\vspace*{0.2cm}
\caption[]{Time development of a tunneling Gaussian wave 
packet in resonant tunneling. The electric field strength
$|E_y|^2$ is shown. The dimensions of the wave guide are the
same as in Fig.~\ref{fig-4}. The spectral function of the
incoming wave packet centered at $\omega_0=\nmbr{18.97}\,\eh{GHz}$
with $\sigma_\omega=\nmbr{0.015}\,\eh{GHz}$ extends over more
than one resonance (see Fig.~\ref{fig-4}), so that multiple
reflection inside the barrier is observed. Times are given in
picoseconds.} 
\label{fig-6}
\end{figure*}

\begin{figure*}
\unitlength=1mm
\noindent
\begin{picture}(162,73)
\put(\picdist,0){\includegraphics{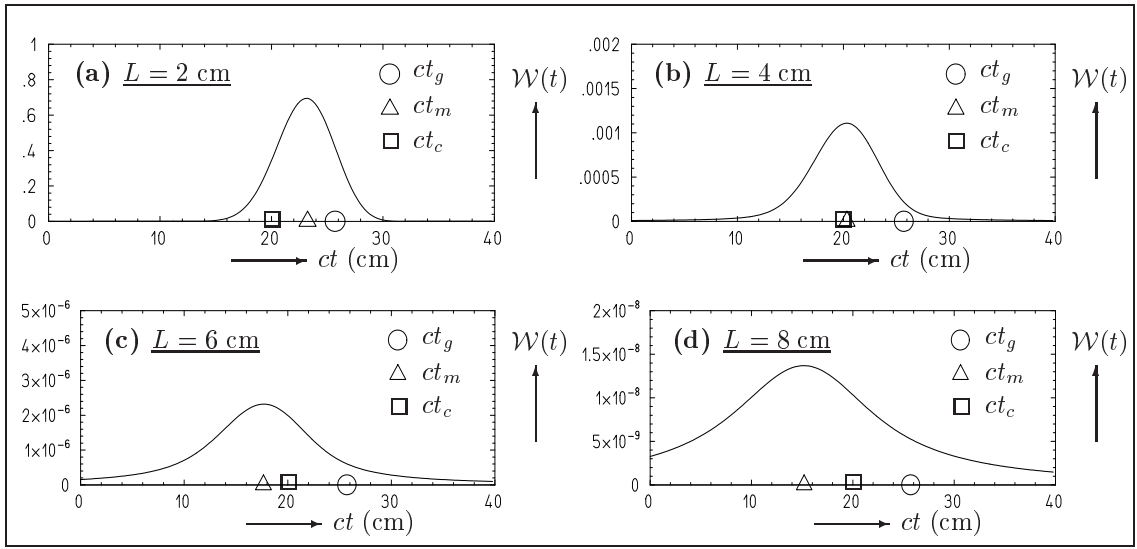}}
\end{picture}
\vspace*{0.2cm}
\caption[]{Example of the superluminal tunneling 
of the maximum of a tunneling Gaussian wave packet. The
longitudinal energy density ${\cal W} (z,t)$ is shown in the
time domain at a fixed position behind the barrier $\Delta
z=\nmbr{20}\,\eh{cm}$ away from the position of the center of the
wave packet at $t = 0$. The dimensions of the symmetric wave
guide are $a=\nmbr{1}\,\eh{cm}$ and $a'=\nmbr{0.5}\,\eh{cm}$ for
different barrier lengths $L$ (see Fig.~\ref{fig-2}). The
incident wave packet is of the type ${\bf H}_{10}$ with
frequencies centered at $\omega_0=\nmbr{15}\,\eh{GHz}$ and
spectral width $\sigma_\omega=\nmbr{0.6}\,\eh{GHz}$. Thus,
evanescent tunneling occurs. The box, the triangle, and the
circle correspond to the arrival times of the pulse maximum in
vacuum, in a wave guide without barrier, and in the wave guide
with a barrier of length $L$, respectively.  If the barrier is
long enough the tunneling velocity of the pulse maximum becomes
superluminal (case {\bf (c)} and {\bf (d)}).} 
\label{fig-7}
\end{figure*}

\begin{figure*}
\unitlength=1mm
\noindent
\begin{picture}(162,62)
\put(\picdist,0){\includegraphics{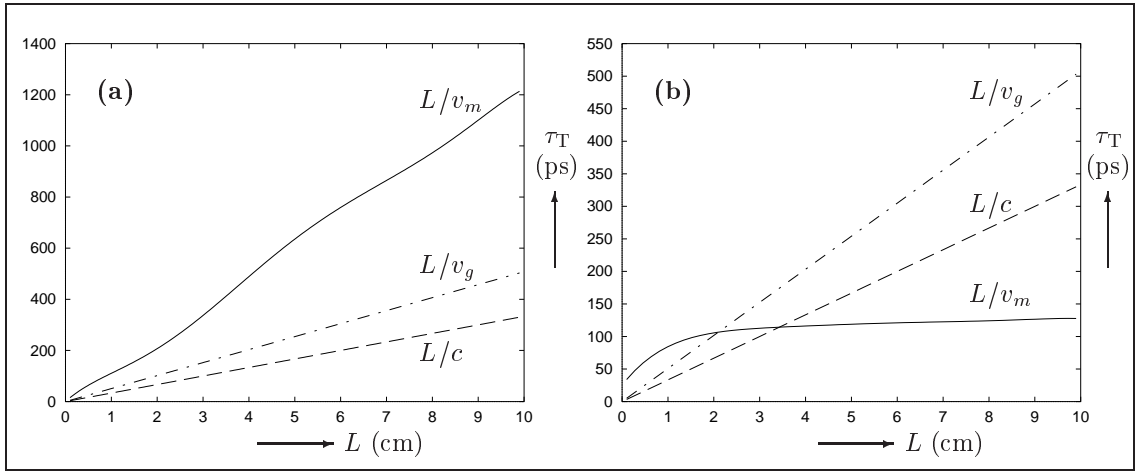}}
\end{picture}
\vspace*{0.2cm}
\caption[]{Tunneling times of the pulse maximum and Hartman 
effect. The transmission time of the pulse maximum (solid line),
the time intervals corresponding to the free motion of the pulse
maximum in vacuum (dashed line) and in the wave guide without
barrier (broken line) for a Kaiser--Bessel wave packet are
plotted as functions of the barrier length $L$. In both cases
the incoming wave packet is a superposition of $N = 800$
frequencies in the interval $\nmbr{51.52}\,{\rm GHz}
\leq\omega\leq \nmbr{57.81}\,{\rm GHz}$ with the spectral
distribution (\ref{eq-22}) for $\alpha = 1$. {\bf (a)}
Non-evanescent tunneling. The cross sections of the wave guide
are $a=\nmbr{22.86}\,\eh{mm}$ and $a' = \nmbr{18}\,\eh{mm}$. 
The behavior of the tunneling times is ``normal''.  {\bf (b)}
Evanescent tunneling. Here, the same tunneling situation as in
\cite{bi-1} is computed, i.e., the two cross sections of the
wave guide are $a = 22.86\,\eh{mm}$ and $a' = 15.8\,\eh{mm}$. 
The transmission time $\tau_{\srr T} = L/v_m$ for long barriers
remains nearly constant and, thus, implies superluminal 
tunneling for the pulse maximum.} 
\label{fig-8}
\end{figure*}

\begin{figure*}
\unitlength=1mm
\noindent
\begin{picture}(85,159)
\put(0,0){\includegraphics{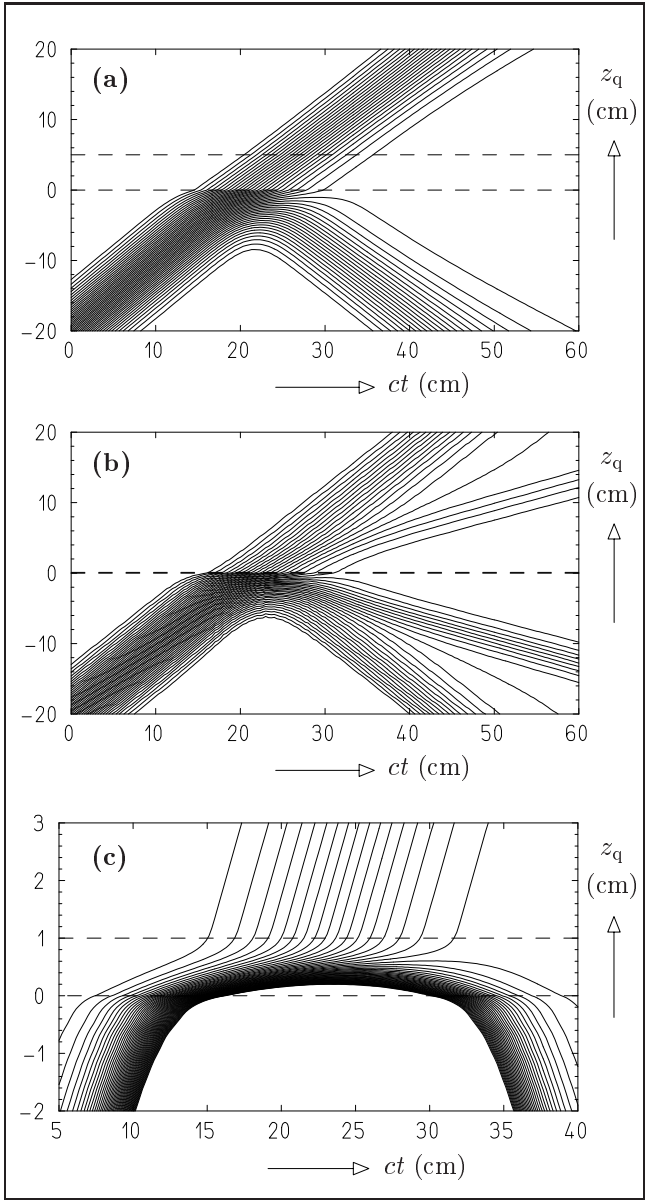}}
\put(\boxdist,0){\makebox(\boxwidth,159)[b]
{\begin{minipage}[b]{\boxwidth mm}\protect\caption[]{Quantile 
trajectories. {\bf (a)} Non-evanescent tunneling. Trajectories 
for $\nmbr{0.05}\leq P \leq \nmbr{0.95}$ in steps of $\Delta P =
\nmbr{0.02}$ are shown. The dimensions of the symmetric wave
guide (see Fig.~\ref{fig-2}) are $a = \nmbr{1}\,\eh{cm}$, $a' =
\nmbr{0.3}\,\eh{cm}$, and $L = \nmbr{5}\,\eh{cm}$. The incoming
Gaussian wave packet is of the type ${\bf H}_{10}$ with
$\omega_0 = \nmbr{60}\,\eh{GHz}$ and $\sigma_\omega =
\nmbr{0.3}\,\eh{GHz}$. {\bf (b)} Evanescent tunneling with short
barrier. The cross sections $a$ and $a'$ are the same as in {\bf
(a)}. The length of the barrier is $L = \nmbr{0.1}\,\eh{cm}$.
The incoming wave packet with $\omega_0 = \nmbr{30}\,\eh{GHz}$
and $\sigma_\omega=\nmbr{0.3}\,\eh{GHz}$ is evanescent in the
region inside the barrier. {\bf (c)} Evanescent tunneling with
long barrier. The length of the barrier is chosen to
$L=\nmbr{1}\,\eh{cm}$. The incoming wave packet has $\omega_0 =
\nmbr{22.5}\,\eh{GHz}$ and $\sigma_\omega =
\nmbr{0.3}\,\eh{GHz}$.  Because of the strong reflection,
trajectories for $10^{-3}\leq P \leq \nmbr{2.5}\times 10^{-2}$
in steps of $\Delta P = \nmbr{5}\times 10^{-4}$ are computed.
Also in this case no superluminal tunneling is observed.}
\end{minipage}}}
\end{picture}
\label{fig-9}
\end{figure*}

\begin{figure*}
\unitlength=1mm
\noindent
\begin{picture}(85,102)
\put(0,0){\includegraphics{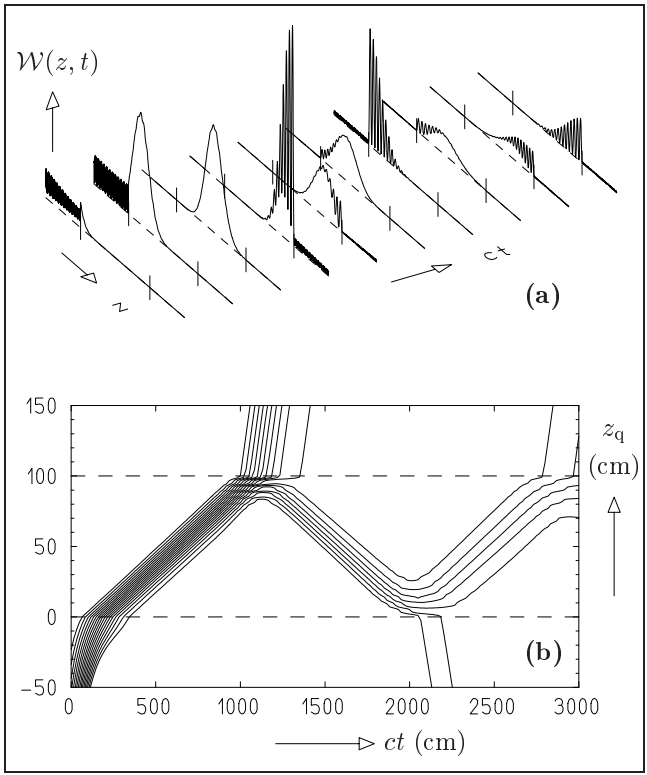}}
\put(\boxdist,0){\makebox(\boxwidth,159)[b]
{\begin{minipage}[b]{\boxwidth mm}\protect\caption[]{Quantile 
trajectories in resonant tunneling. The dimensions of the
symmetric wave guide are $a = \nmbr{1}\,\eh{cm}$, $a' =
\nmbr{0.5} \,\eh{cm}$, and $L = \nmbr{100}\,\eh{cm}$. The
incoming wave is a Gaussian wave packet of the type ${\bf
H}_{10}$ with $\omega_0 = \nmbr{18.97}\, \eh{GHz}$ and
$\sigma_\omega = \nmbr{0.015}\,\eh{GHz}$. {\bf (a)} Time
development of the longitudinal energy density ${\cal W}(z,t)$.
{\bf (b)} Quantile trajectories for $\nmbr{0.2}\leq P \leq
\nmbr{0.625}$ in steps of $\Delta P = \nmbr{0.025}$. The
oscillating behavior of the quantile trajectories reflects the
time development shown in {\bf (a)}.}
\end{minipage}}}
\end{picture}
\label{fig-10}
\end{figure*}

\end{document}